# Baseline wander removal methods for ECG signals: A comparative study

*Francisco Perdigón Romero [1], Liset Vázquez Romaguera [1],*
*Carlos Román Vázquez-Seisdedos [2], Cícero Ferreira Fernandes Costa Filho [3],*
*Marly Guimarães Fernandes Costa [3], João Evangelista Neto [3,4]*

[1] École Polytechnique de Montréal, Canada
[2] Center for Neuroscience Studies, Images and Signals Processing, Universidad de Oriente, Cuba
[3] Center for R&D in Electronic and Information Technology, Universidade Federal do Amazonas, Brasil
[4] Universidade Estadual do Amazonas, Brasil

**ABSTRACT**

Cardiovascular diseases are the leading cause of death worldwide, accounting for 17.3 million deaths per year. The electrocardiogram (ECG) is a non-invasive technique widely used for the detection of cardiac diseases. To increase diagnostic sensitivity, ECG is acquired during exercise stress tests or in an ambulatory way. Under these acquisition conditions, the ECG is strongly affected by some types of noise, mainly by baseline wander (BLW). In this work were implemented nine methods widely used for the elimination of BLW, which are: interpolation using cubic splines, FIR filter, IIR filter, least mean square adaptive filtering, moving-average filter, independent component analysis, interpolation and successive subtraction of median values in RR interval, empirical mode decomposition, and wavelet filtering. For the quantitative evaluation, the following similarity metrics were used: absolute maximum distance, the sum of squares of distances and percentage root-mean-square difference. Several experiments were performed using synthetic ECG signals generated by ECGSYM software, real ECG signals from QT Database, artificial BLW generated by the software and real BLW from the Noise Stress Test Database. The best results were obtained by the method based on FIR highpass filter with a cut-off frequency of 0.67 Hz.

**1. INTRODUCTION**

Cardiovascular diseases are the leading cause of sudden cardiac death in the world [1]. Therefore, there is a great demand for specialized medical services such as diagnostic tools for the study and treatment of patients with these diseases.

The electrocardiogram (ECG) is the simplest non-invasive technique and, at the same time, the most used in the diagnosis of heart diseases. The ECG signal is the electrical manifestation of the heartbeat over time and can be recorded under various conditions in order to detect different types of abnormalities in the heart. These conditions of registration are: at rest, in ambulatory conditions or during stress tests [2].

The resting ECG mode records the electrical activity of the heart while the subject is in the dorsal decubitus position. Since these recordings have a short duration, the detection of events that vary over long periods of time or related to certain physical activities are not detected. Examples of these events are the identification of left ventricular hypertrophy and left atrial overload, among others.

The ambulatory ECG is acquired while the patient performs his daily activities, for a period of 24 hours or more, so the probability of registering one or more of the events mentioned above is greater. In the physical exertion test, the patient performs for a short period of time (5 - 15 min), a controlled physical effort on a bicycle or on an electric treadmill. In this circumstance, it is possible to assess through the ECG the functional capacity of the heart of the subject that is being evaluated, being able to detect possible arrhythmias, the occurrence of myocardial ischemia, among others.

ECG records acquired on an ambulatory way or in stress tests are heavily contaminated with noise, given the conditions in which they are acquired [3]. One of the main sources of noise in these acquisition conditions is the baseline wander (BLW).

BLW are low frequency noise (0.05 - 3 Hz in stress tests) [3,4]. They are mainly due to movement during breathing, patient movements, poor contact between electrode cables and ECG recording equipment, inadequate skin preparation where the electrode is placed and dirty electrodes.

BLW severely limit the usefulness of ECG records, especially when they are acquired in an ambulatory way or during physical exertion tests. Most of the algorithms for detection of cardiac pathologies or events associated with the activity of the heart present a malfunction in the presence of BLW, which is why the elimination of this kind of noise is necessary to guarantee a better clinical evaluation.

There are previous works in which the authors compiled and explained several methods for the elimination of BLW [5-9]. It is important to emphasize that the previously mentioned works do not include the implementation and comparison of all the compiled methods using different metrics.

Next, we will mention some works in which the authors compared some methods for the BLW elimination. Kumar et al. presented a comparison of methods based on classic digital filtering. The metrics used were signal to noise ratio (SNR) and mean square error (MSE) [6]. Chaudhary et al. performed a comparison between different wavelets and thresholds in order to find the combination that generates the least distortion in the ECG signal when eliminating the BLW. The metrics used in this work were MSE and SNR [7].

On the other hand, Narwaria et al. made a comparison between different types of IIR filters (Butterworth, Elliptic, Chebyshev I and Chebyshev II). The metrics used were SNR and a qualitative evaluation of the modification of the ECG waveforms [8].

Another work that must be taken into account is that of Joshi et al., which compared methods based on Kalman filter, moving-average filter, and cubic splines. The metrics used were MSE and the standard deviation [9].

Although some methods are compared in the aforementioned works, there are other modern and more complex that are not considered, such as those based on empirical mode decomposition, independent component analysis, cascade adaptive filter and based on interpolation and successive subtraction of median values in RR intervals. On the other hand in most of these works the metrics used were the MSE or the SNR. These metrics are not the most adequate to quantify the level of distortion suffered by the ECG signal during the BLW elimination. The SNR is a measure of quality, not of similarity. Therefore, when used in the context of BLW elimination, this can show good results even though the ECG signal presents considerable distortions. On the other hand, the MSE, despite being a measure of similarity, can mask large distortions of the signal as it performs an average of the error [10,11]. Considering this, in our comparative study we will use similarity metrics based on the distance that does not perform average.

Despite the fact that the ECG is one of the most researched biosignals, there is still much to be done regarding the BLW. In the literature, there are many works that deal with the elimination of DLB [12-20].

Most of the referenced works use their own database and different performance metrics. This makes it impossible to perform an impartial comparison between the most commonly used methods for BLW removal.

In the present work, a comparative study of nine of the most cited methods for BLW removal was carried out. The methods were implemented using MATLAB R2014a. A set of common signals was used for all the methods. We presented experiments where real BLWs were added to the ECG signal. This represents a novelty since, to the best of our knowledge, has not been made in similar works. Each method was evaluated through similarity metrics. The implementation of these methods and metrics used in this work are available on the Github platform as open source[1]. Therefore, this will facilitate that other researchers compare new methods for BLW removal using a common framework.

## 2. BIBLIOGRAPHIC REVIEW

In this session, the 9 techniques chosen for the comparative study will be presented. The way in which they were selected will also be discussed. The methods selected to perform this comparative study were selected taking into account the relevance shown by the search engines used and the number of citations [21]. In Table 1 the selected methods are presented with their citations scores.

The databases used were: IEEE Xplore (IE), Web of Science (WoS) and Google Scholar (GS).

---

[1] https://github.com/fperdigon/ECG-BaseLineWander-Removal-Methods/

Table 1: Citation scores of the implemented methods

| BLW removal technique | Number of citations | | |
|---|---|---|---|
| | IE | WoS | GS |
| Finite impulse response (FIR) high pass filter [12] | 138 | 131 | 300 |
| Infinite impulse response (IIR) high pass filter [13] | - | - | 41 |
| Cubic splines [14] | - | 131 | 240 |
| Two cascades least mean square (LMS) adaptive filters [15] | 21 | 1 | 15 |
| Moving-average filter [16] | 1 | 5 | 24 |
| Independent components analysis (ICA) [17] | - | 1 | 15 |
| Interpolation and successive subtraction of median values in RR intervals (ISSM) [18] | 26 | 5 | 70 |
| Empirical mode decomposition (EMD) [19] | - | 124 | 304 |
| Wavelet filter [20] | - | - | 40 |

**The method based on finite impulse response high pass filter:** Currently among the most used methods for BLW removal are classic digital filters. In the work carried out by Van Alsté and Schilder [12], the authors calculated and implemented an FIR high pass filter using a size 28 Kaiser window with 51 coefficients. The analysis of the distortion in the signal was qualitative.

**The method based on the infinite impulse response high pass filter**: This type of filter can reach an accentuated transition region with fewer coefficients than an FIR filter. However, IIR filters have a non-linear phase response that distorts the ECG signal. Bi-directional filters are used to avoid this distortion. The IIR filter implemented Pottala and collaborators [6] used a small window so that it can be used in real-time applications. The analysis of the distortion in the signal was qualitative.

**The method based on interpolation using cubic splines**: Meyer and Keizer [14] estimated a point in the isoelectric PR-segment of each beat. For this, they used the point R as a reference and estimated a point 66 ms before the detected R-point with normally point to the PR isoelectric point. Then, based on those points, they performed an interpolation using cubic splines to estimate the BLW and finally subtracted the estimated BLW from the ECG signal. The cubic splines are differentiable curves that are formed from polynomials of degree 3. The analysis of the distortion in the signal was qualitative.

**The method based on cascade least mean square filter**: Laguna and collaborators proposed a 2 stage cascade LMS filter for the elimination of BLW [15]. In the 2 stages, adaptive LMS filters were used. In the first stage, a high-pass Notch filter was used as input of the adaptive filter. The second stage was formed by a correlated filter of adaptive impulse. The analysis of the distortion in the signal was qualitative.

**The method based on the moving-average filter**: Canan and collaborators used a moving-average filter to eliminate BLW [16], which behaves like a low pass filter but works in the time domain. In this technique, an estimated BLW was first extracted. Then it was subtracted from the original signal. It was necessary to fill with zeros at the beginning and at the end of the estimated BLW due to the loss of samples that occurs when using a sliding window. The authors affirmed that this method allowed eliminating the BLW in a simple way, without having to calculate the coefficients of the FIR or IIR filters. The analysis of the distortion in the signal was qualitative.

**The method based on independent components analysis**: In that work, the authors used the independent component analysis technique to eliminate the BLW [17]. To use this technique it is necessary that the signal has multiple channels. When only one channel is available, as is the case of ECG signals used by the authors, the method described by J.Lee [22] can be used. In this approach, versions of the same signal with different delays work as a set of multiple channels. Barati and Ayatollahi generated 60 channels from a single ECG signal using versions with delays between 11 and 20 samples, which form the input set called mixed element matrix. They used the FastICA algorithm. The independent components corresponding to BLW were detected automatically using the kurtosis. The kurtosis is the fourth statistical moment and, according to Delorme's research [23], when it is negative corresponds to BLW. Finally, the components that were detected as BLW are set to zero and the original matrix is recomposed to extract the original ECG. The analysis of the distortion in the signal was quantitative and the metrics used were: sensitivity, specificity, and accuracy.

**The method based on interpolation and subtraction of values of the median of the signal in the RR intervals**: Chouhan and Mehta proposed a method based on the analysis of RR intervals [18] for the BLW removal. In this approach, the median value of the ECG signal in each RR interval is calculated and subtracted from the values of each sample in the RR interval under analysis. The previous process is repeated until the median values of all RR intervals have a small difference defined by a threshold. The authors did not use any metric to evaluate the performance of this method. The analysis of the distortion in the signal was qualitative.

**The method based on the empirical mode decomposition**: Blanco-Velazco and collaborators presented an EMD based technique to eliminate the BLW [19]. First, the ECG was decomposed in several intrinsic mode functions (IMF). According to the authors, the higher order IMF contains the information referring to the BLW but they do not advise the direct elimination of them but they emphasize that they must be filtered using a type IIR low pass filter. Then, an estimated BLW was obtained which is subtracted from the original ECG signal. The analysis of the distortion in the signal was quantitative, through the mean square error.

**Method based on Wavelet filter**: One of the pioneering works that used Wavelet filter for the elimination of BLW was presented by Mozaffary and Tinati in 2005 [20]. Later, in 2015, Chaudhary and collaborators made a comparative study [7] to determine which of the wavelet families (Haar, Db5, Coif3, Bior3.1, Db4, Sym8, Sym10, Bior6.8, Db6, Coif4) and what type of threshold (Hard, Soft, Semi-soft, Stein) should be used. The combination that yielded the best results was Wavelet Sym10 with a Semi-soft threshold. The analysis of the distortion in the signal was quantitative. To calculate the distortion in the ECG signal, the mean square error metric was used.

## 3. MATERIALS

For the experiments, records belonging to the QT Database [24] of Physionet [25] were used. This database contains 105 15-min two-lead ECG recordings sampled at 250 Hz. The records chosen for the experiments were 3: *sel100m*, *sel102m*, and *sel116m*. For the experiments, the first 5 min of channel 1 of each selected record was used. These three records were chosen due they present a practically zero BLW, which is very convenient. Figure 1 shows a 5 s segment of the *sel100m* signal.

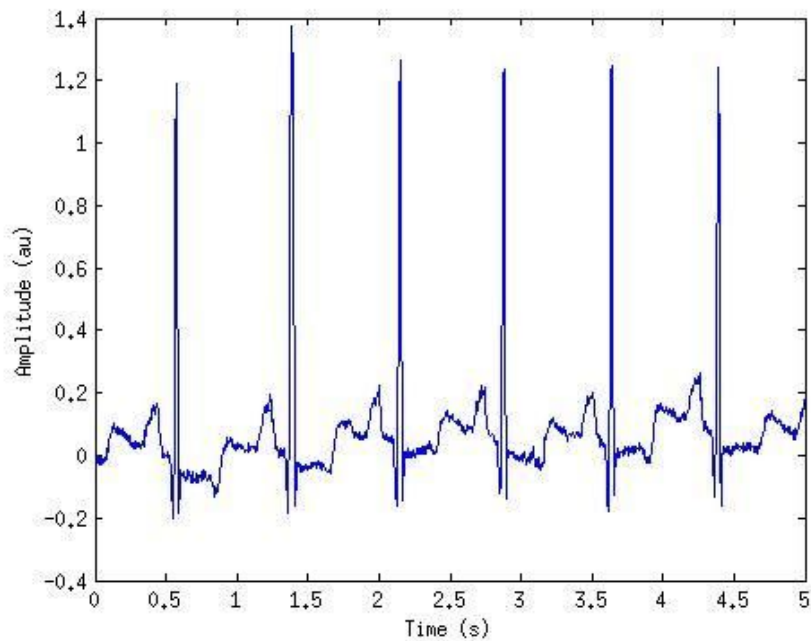

**Figure 1. A 5 s segment of the signal *sel100m*.**

The artificial ECG signals were generated using the ECGSYM software [26] of Physionet [25] which allows configuring several ECG parameters such as the heart rate, sampling frequency of the resulting signal, number of beats, the morphology of the ECG waves (P, Q, R, S and T), amplitude and duration parameters, etc. For the experiments, a sampling frequency was set to 360 Hz and the duration of the signal to 5 min. Signals with different heart rates were generated, the first at 70 beats per minute (bpm) which is considered normal and the other at 120 bpm which simulates a person's heart rate during the performance of a physical exertion test. Figure 2 shows a 5 s segment of an artificial ECG signal of 70 bpm generated by the ECGSYM software.

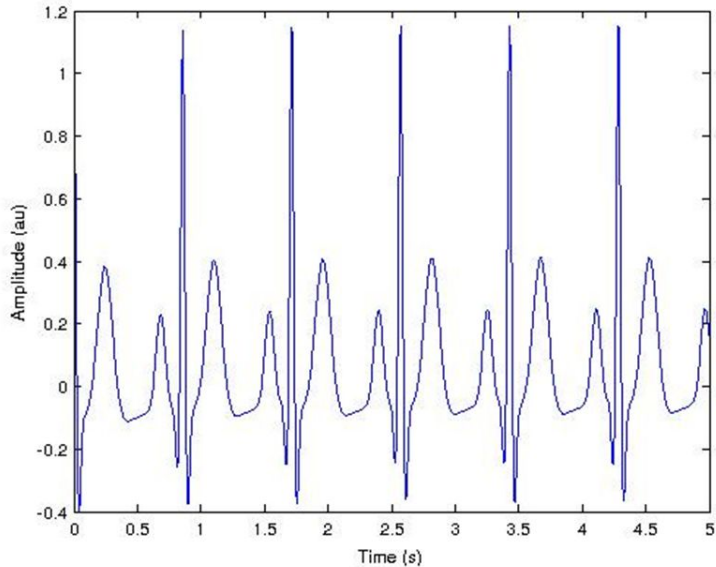

**Figure 2. A 5 s segment of artificial ECG signal of 70 bpm generated by the ECGSYM software.**

In the experiments, real baseline drifts were used. These were obtained from the MIT-BIH Noise Stress Test Database (NSTDB) [27] of Physionet [25] which contains 12 ECG records of 30 min and 3 records of 30 min of typical noises present in the ECG during the stress tests: baseline wander produced by the patient's breathing, artifacts generated by the movement of the electrodes and electromyographic noise. In this database, the ECG records are contaminated randomly with the noise present in the 3 noise channels. The sampling frequency is 360 Hz. Noisy records were acquired while volunteers performed physical stress tests through electrodes placed at limb positions where the ECG cannot be captured. For the experiments, we used the records that contained the baseline wander generated by respiration and the recording with artifacts generated by the movement of the electrodes, see Figure 3.

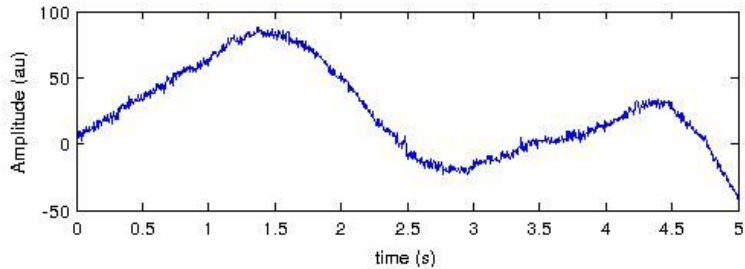

(a)

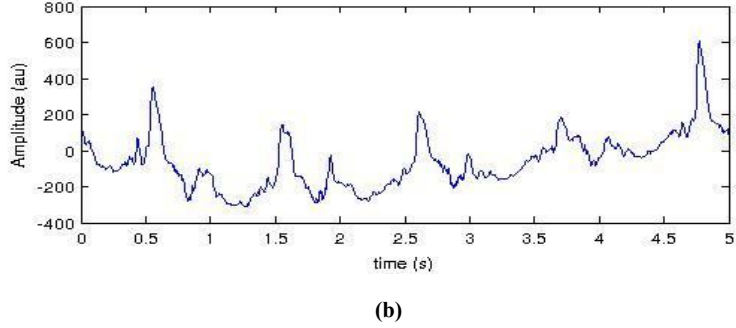

(b)

**Figure 3. A 5 s segment of the NSTDB (a) record "bw" channel 1, BLW generated by the breath (b) record "em" channel 1 artifact generated by the movement of the electrodes.**

For the implementation of the ICA-based method, the FastICA-CIS toolbox developed at the Laboratory of Information and Computer Science at the Helsinki University of Technology, Finland [28] was used. For the implementation of the EMD-based method the EMB Bivariate Empirical Mode Decomposition Approach toolbox, developed at Laboratoire de Physique, Ecole Normale Superieure de Lyon, France [29] was used. For the experiments, the MATLAB 2014a software was used in a Core I3 computer with 6GB of RAM memory with Ubuntu GNU/Linux 14.04 operating system.

## 3. METHODS

To evaluate the distortion of the ECG signal in a quantitative way, an analysis of several similarity metrics was made, these are:
- Metrics based on statistics: Mean square error (MSE), Cross-correlation (CC), Signal noise ratio.
- Metrics based on distance: Absolute maximum distance (MAD), sum of the square of the distances (SSD), percentage root-mean-square difference (PRD).
As was mentioned before, according to studies published in [10,11], distance-based metrics are the most suitable for the evaluation of signal similarity. Below we explain the selected metrics.

**Absolute maximum distance**: It is one of the most common similarity metrics used to determine the quality of ECG signals after performing a compression process [11] and can be defined by the following expression:

$$MAD(s1, s2) = \max |s2(m) - s1(m)| \quad 1 \leq m \leq r \qquad (1)$$

where *s1* and *s2* are the signals to be compared, *m* is the number of the current sample of the signals and *r* is the maximum number of samples of the *s1* and *s2* signals.

**Sum of the square of the distances**: It is another of the similarity metrics used to evaluate the distortion between signals. It allows to measure the accumulated error and gives an idea of how different the signals are in all their extension [11].

$$SSD(s1, s2) = \sum_{m=1}^{r} (s2(m) - s1(m))^2 \quad 1 \leq m \leq r \qquad (2)$$

Where *s1* and *s2* are the signals to be compared, *m* is the number of the current sample of the signals and *r* is the maximum number of samples of the *s1* and *s2* signals.

**Percentage root-mean-square difference**: It is a widely used similarity metric that, despite using the mean which is a measure of central tendency, is based mainly on distance [11].

$$PRD(s1, s2) = \sqrt{\frac{\sum_{m=1}^{r}(s2(m) - s1(m))^2}{\sum_{m=1}^{r}(s2(m) - \overline{s1})^2}} \times 100\% \quad 1 \leq m \leq r \quad (3)$$

Where *s1* and *s2* are the signals to be compared, *m* is the number of the current sample of the signals and *r* is the maximum number of samples of the *s1* and s2 signals.

To carry out the comparative study of the nine methods, these metrics were evaluated in ECG signals artificially generated by the ECGSYM software [26] and in real signals from the QT Database [24]. These signals were contaminated with artificial BLW (sine signals at different frequencies) and by real BLW from the MIT-BIH Noise Stress Test Database [27]. After applying the nine methods for the elimination of the BLW, the three selected metrics were calculated. Because the signals of the QT Database have a sampling frequency of 250 Hz, the noise of the NSTDB was resampled to this frequency since it was originally sampled at 360 Hz. Table 2 shows the characteristics of the experiments performed.

Table 2. Experiments description

| Elements | Value |
| --- | --- |
| Signals (duration 5 min) | Artificial (ECGSYM): 120 bpm<br>Real (QT Database): *sel100m*, *sel103m*, *sel116m*.<br>First 5 min. |
| BLW type (duration 5 min) | Artificial (sinusoidal signal): 0.60 Hz.<br>Real (NSTDB). First 5 min. |
| Methods for the elimination of BLW | Interpolation using cubic splines, FIR filter, IIR filter, LMS adaptive filter, moving-average filter, ICA, successive medial interpolation and subtraction, EMD, and wavelet filters. |
| Similarity metrics | MAD, SSD, PRD. |

During the experiments, the methods using high-pass filtering were configured with the cut-off frequency of 0.67 Hz, which is the value recommended by the AHA [4]. To standardize the level of contamination with the different BLW, a value of 0.5 a.u. of MAD metric was set with respect to the original signal.

## 4. RESULTS AND DISCUSSION

In this section, the results of the experiments will be discussed. The reader interested in seeing other related experiments can consult the reference [21].

In addition to the numerical result in the columns, in the lasts columns of the table were placed the positions that each of the methods reached according to each of the metrics. In the legend of the tables are important information regarding the signal (artificial or real), the heart rate in the case of artificial signals (hr), type of noise (artificial or real) and cut-off frequency (fc) of the methods that use filtering.

For the qualitative analysis of the results, images of the signals will be shown at the point where the MAD metric obtained the highest (worst) results.

**Table 3. Performance of the implemented methods evaluated in artificial ECG (hr = 120 bpm), artificial BLW (sine wave 0.60 Hz) and cut-off frequency, fc = 0.67 Hz.**

| Method | Metric value | | | Classification order | | |
|---|---|---|---|---|---|---|
| | MAD | SSD | PRD | MAD | SSD | PRD |
| **Splines** | 0,16 | 814,38 | 36,37 | 1º | 1º | 1º |
| **FIR** | 0,22 | 3565,87 | 79,98 | 2º | 3º | 2º |
| **IIR** | 0,72 | 15949,51 | 169,15 | 6º | 7º | 7º |
| **AF** | 1,16 | 3520,16 | 84,49 | 8º | 2º | 5º |
| **MAF** | 0,37 | 5108,90 | 91,10 | 4º | 6º | 6º |
| **ICA** | 21,73 | 1721755,77 | 1672,32 | 9º | 9º | 9º |
| **ISSM** | 0,40 | 4326,17 | 83,83 | 5º | 5º | 4º |
| **EMD** | 0,94 | 17744,50 | 169,77 | 7º | 8º | 8º |
| **WT** | 0,24 | 3964,97 | 80,25 | 3º | 4º | 3º |

Splines: Interpolation using cubic splines; FIR: FIR filter; IIR: IIR filter; AF: Adaptive filter; MAF: Moving-average filter; ICA: Independent components analysis; ISSM: Interpolation and subtraction of values of the median of the signal in RR intervals; EMD: Empirical mode decomposition WT: Wavelet Transform filter; MAD: Absolute maximum distance; SSD: Sum of the square of the distances; PRD: Percentage root-mean-square difference.

Table 3 shows that, in the three metrics used, the method based on interpolation using cubic splines showed the best performance. This is due to the combination of 2 factors: the first is that the heart rate is 120 bpm (in 1 s the algorithm detects 2 points for interpolation) and the second is that the noise to be interpolated is a sinusoid, which is a low complexity signal. This combination allows the algorithm to work in almost ideal conditions. Figure 4 shows the performance of the method based on interpolation using cubic splines on artificial ECG, 1 s segment, centered on the point of greatest distortion, according to the MAD metric.

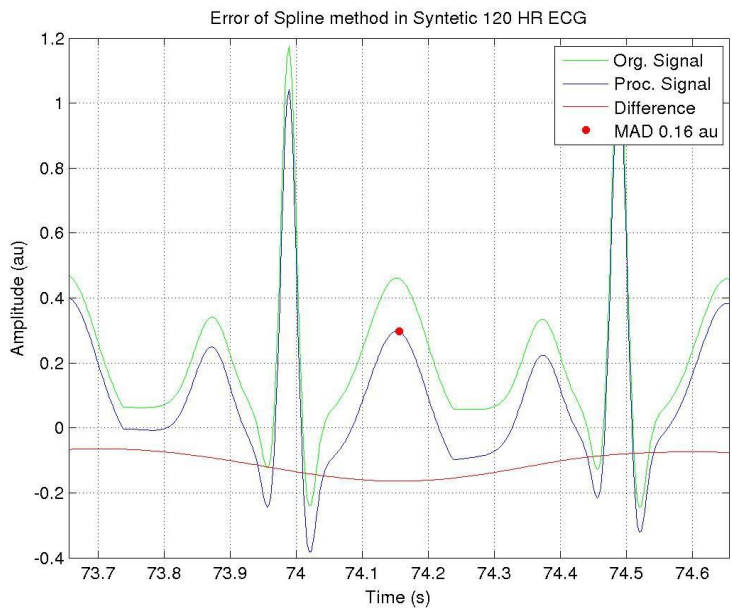

**Figure 4. Performance of the method based on interpolation using cubic splines in artificial ECG. 1 s segment, centered on the point of greatest distortion, according to the MAD metric. The green trace shows the original signal.**

The second-best performance, according to the MAD and PRD metrics, was yielded by the FIR filter method. Although this kind of filter presents a computational cost greater than other digital filters, it is very efficient. The use of many coefficients allows the filter to have a high slope, which contributes to the efficient elimination of noise. In Figure 5 can be observed that there is almost no distortion in the morphology of the signal. Also, the difference between both signals is practically a straight line, which allows us to infer that it is almost a bias value.

According to the MAD and PRD metrics, the third-best performance was reached by the method based on wavelet transform. In Figure 6 can be observed that the distortion in the morphology of the signal is very small.

Interestingly, despite being in $8^{th}$ place, according to the MAD metric and $5^{th}$ place, according to the PRD metric, the method based on the adaptive filter is in $2^{nd}$ place according to the SSD metric. With the help of Figure 7, we can explain the reason for this behavior. The SSD metric measures the accumulated error. This means that, although the error had high values in certain points of the signal, as shown in Figure 7, the sum of these errors is less than the sum of the errors generated by the other methods. It is important to mention that the distortion of the R point shown in Figure 7 is due to the proper functioning of the method since, from all R point, the windows where the filters applied are created. The effect of the distortion that is observed is generated because the windows are not correctly joined by the algorithm.

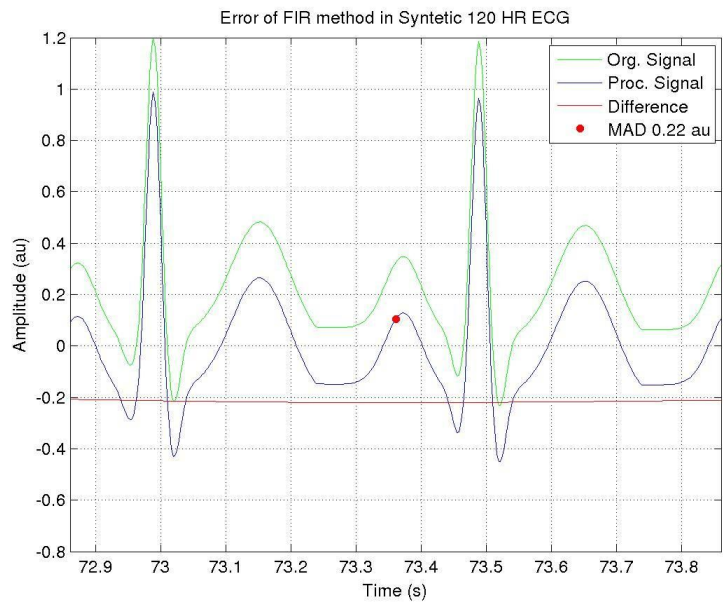

**Figure 5. Performance of the method based on FIR filter on artificial ECG. 1 s segment, centered on the point of greatest distortion, according to the MAD metric. The green trace shows the original signal. The blue trace shows the signal processed by the method. The red trace shows the difference between the 2 analyzed signals.**

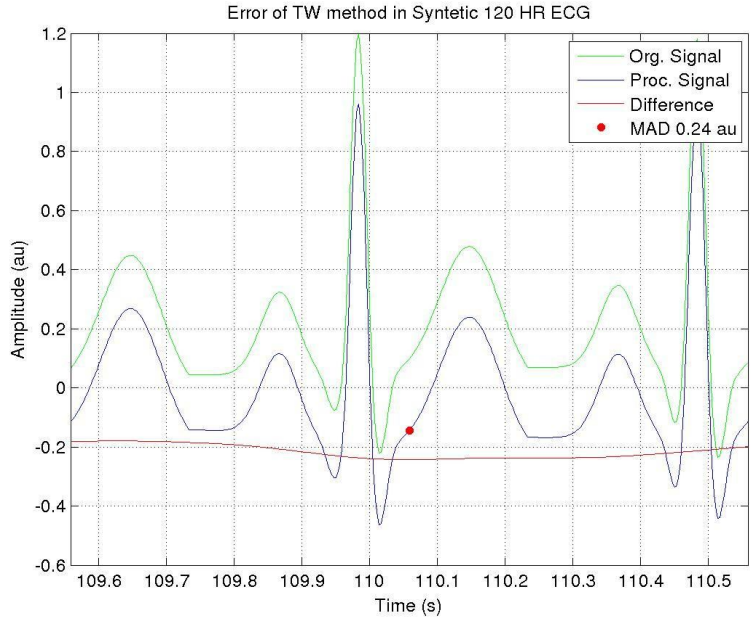

**Figure 6. Performance of the method based on wavelet transform in artificial ECG. 1 s segment, centered on the point of greatest distortion, according to the MAD metric. The green trace shows the original signal. The blue trace shows the signal processed by the method. The red trace shows the difference between the 2 analyzed signals.**

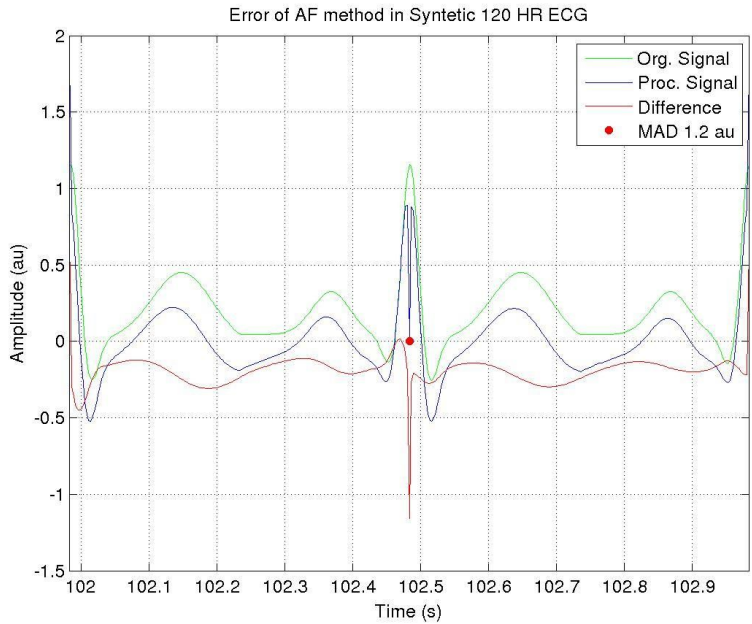

**Figure 7. Performance of the method based on adaptive filter in artificial ECG. 1 s segment, centered on the point of greatest distortion, according to the MAD metric. The green trace shows the original signal. The blue trace shows the signal processed by the method. The red trace shows the difference between the 2 analyzed signals.**

In the last position, according to all the metrics, is the method based on ICA. In Figure 8 can be observed that the output signal is completely distorted. This apparent malfunction of the method can be explained through the theoretical bases of the independent components analysis. In the work of Hyvärinen and Oja [30] was explained that one of the restrictions of ICA is that the components have to be statistically independent or not correlated. This means that the covariance of the signals must be zero. Figure 9 (a) and (b) show a graph of the covariance between the artificial ECG of 120 bpm and the simulated ECG signal, through a sinusoid of 0.60 Hz. As can be observed, covariance presents high values close to half the peak value of the signal and a sinusoidal shape. Given the characteristics of the signal and noise, it is understandable that the ICA-based method presents the obtained performance.

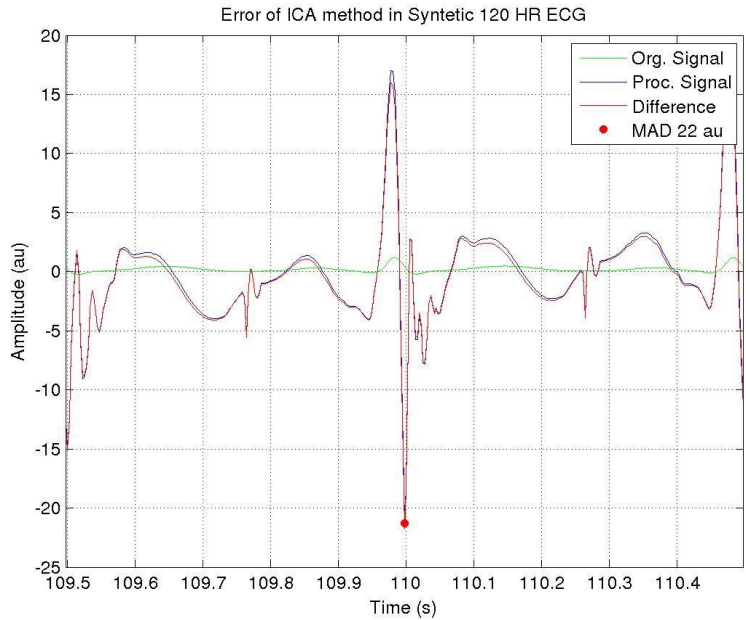

**Figure 8. Performance of the method based on ICA in artificial ECG. 1 s segment, centered on the point of greatest distortion, according to the MAD metric. The green trace shows the original signal. The blue trace shows the signal processed by the method. The red trace shows the difference between the 2 analyzed signals.**

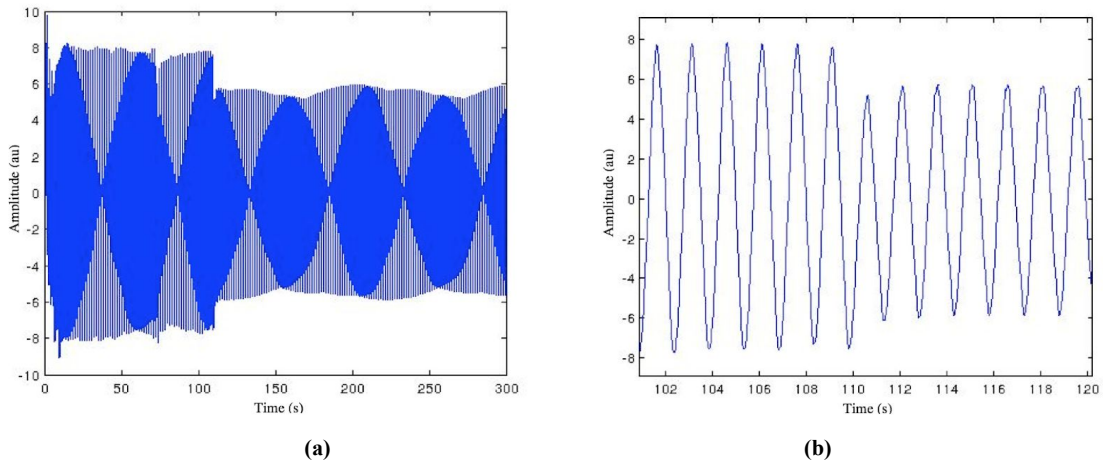

(a)          (b)

**Figure 9. Graphs showing the covariance between the artificial ECG signal with hr = 120 bpm and an artificial BLW simulated through a 0.6 Hz sinusoid. (a) Covariance plot in the 5 min of the 2 signals, (b) Extended graph.**

Table 4 shows the results of the experiments with real signals of the QT Database and the real noise of the NSTDB. These experiments, despite being in a controlled environment, are the most similar to real situations.

**Table 4. Performance of the implemented methods evaluated in real ECG (QT Database), real BLW (NSTDB) and cut-off frequency, fc = 0.67 Hz.**

| Method | Metric value | | | Classification order | | |
|---|---|---|---|---|---|---|
| | MAD | SSD | PRD | MAD | SSD | PRD |
| **Splines** | 0,61 | 3458,01 | 49,88 | 6º | 6º | 7º |
| **FIR** | 0,20 | 152,00 | 16,58 | 1º | 1º | 1º |
| **IIR** | 0,24 | 176,45 | 17,86 | 2º | 2º | 2º |
| **AF** | 2,54 | 4520,01 | 70,523 | 9º | 7º | 8º |
| **MAF** | 0,34 | 274,92 | 18,77 | 3º | 3º | 3º |
| **ICA** | 1,65 | 11877,40 | 94,96 | 8º | 9º | 9º |
| **ISSM** | 0,44 | 1137,49 | 29,64 | 5º | 5º | 5º |
| **EMD** | 0,42 | 306,82 | 20,39 | 4º | 4º | 4º |
| **WT** | 1,29 | 4724,13 | 42,02 | 7º | 8º | 6º |

Splines: Interpolation using cubic splines; FIR: FIR filter; IIR: IIR filter; AF: Adaptive filter; MAF: Moving-average filter; ICA: Independent components analysis; ISSM: Interpolation and subtraction of values of the median of the signal in RR intervals; EMD: Empirical mode decomposition WT: Wavelet filter; MAD: Absolute maximum distance; SSD: Sum of the square of the distances; PRD: Percentage root-mean-square difference.

The best performance, according to the three metrics used, was yielded by the method based on the FIR filter. As previously explained, this kind of filter is more computationally expensive than other digital filters. However, given the advances of the microprocessors used in current embedded systems, it can be implemented in such systems as well as in those based on personal computers. Figure 10 shows the performance of the method based on FIR filter on real ECG contaminated with real noise, in a segment of 1 s, centered on the point of greatest distortion, according to the MAD metric.

The second-best performance, according to the metrics used, was yielded by the method based on IIR filter. This kind of digital filter, despite being less efficient than the FIR filter, has shown a good performance. Figure 10 shows the performance of the method based on FIR filter on real ECG contaminated with real noise, in a segment of 1 s, centered on the point of greatest distortion, according to the MAD metric.

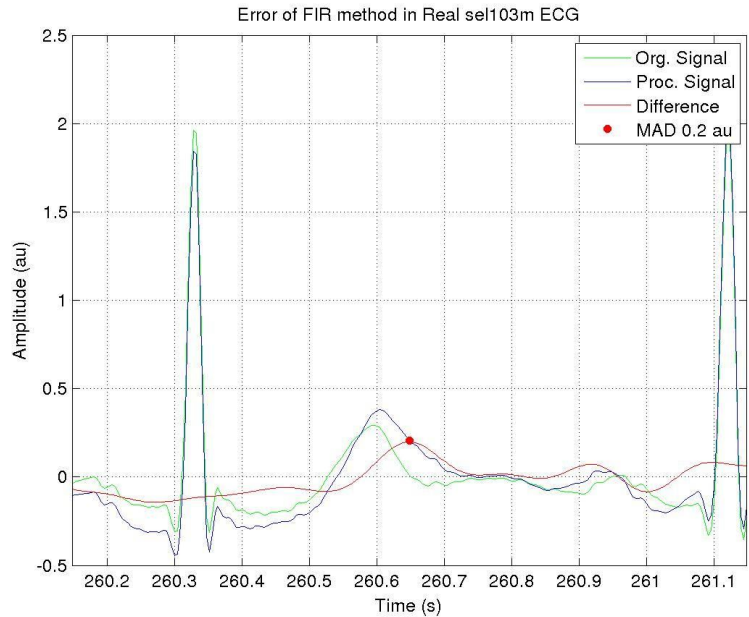

**Figure 10.** Performance of the method based on FIR filter on real ECG contaminated with real noise. 1 s segment, centered on the point of greatest distortion, according to the MAD metric. The green trace shows the original signal. The blue trace shows the signal processed by the method. The red trace shows the difference between the 2 analyzed signals.

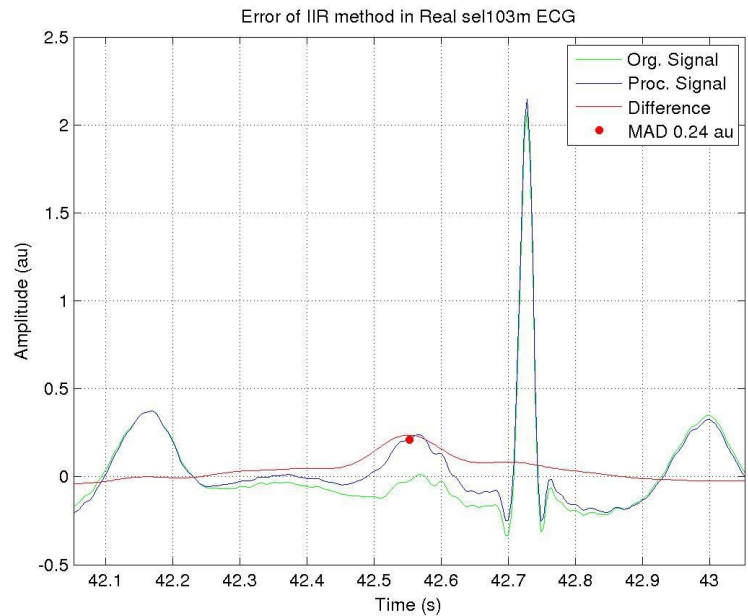

**Figure 11.** Performance of the method based on IIR filter on real ECG contaminated with real noise. 1 s segment, centered on the point of greatest distortion, according to the MAD metric. The green trace shows the original signal. The blue trace shows the signal processed by the method. The red trace shows the difference between the 2 analyzed signals.

The third best performance, according to the three metrics used, was reached by the method based on the moving average filter, which is a kind of filter that works in the time domain. Despite the place obtained, in Figure 12 can be observed that the point where the greatest distortion is found, according to the MAD metric, is on the P wave of the ECG signal and it is a considerable distortion.

According to the MAD metric, the method based on the adaptive filter shows the worst performance. In Figure 13 can be observed that the greatest distortion is close to point R. As was explained before this is due to the proper functioning of the method.

According to the accumulated error metrics SSD and PRD, the worst performance was obtained by the ICA-based method. The possible reasons are those mentioned in the results of the experiments with artificial ECG and noise signals. The BLW due to respiration is very similar to sinusoids and the P and T waves of the ECG are very similar to the positive semi-cycles of a sinusoid. This makes the covariance between the BLW and the ECG high and causes the malfunction of this method. Figure 14 shows the performance of the ICA-based method in real ECG contaminated with real noise, in a window of 1 s, centered on the point of greatest distortion, according to the MAD metric.

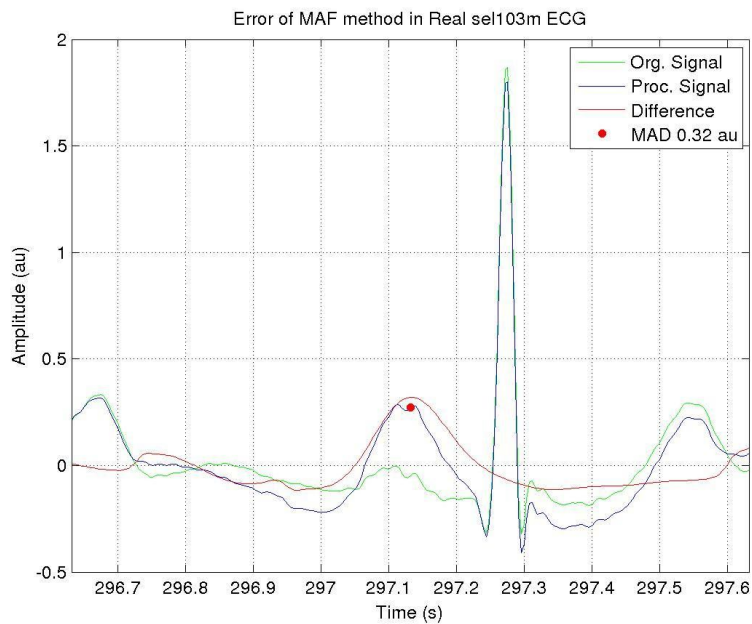

Figure 12. Performance of the method based on moving-average filter on real ECG contaminated with real noise. 1 s segment, centered on the point of greatest distortion, according to the MAD metric. The green trace shows the original signal. The blue trace shows the signal processed by the method. The red trace shows the difference between the 2 analyzed signals.

The difference in the results obtained by the methods in Table 3 and Table 4 is mainly due to the nature of the BLW used in each experiment. The most relevant case is that the method based on cubic splines in Table 3 is in the first position and in Table 4 it goes to the sixth position. As previously analyzed, this is due to the fact that in the experiment analyzed in Table 3 the BLW is modeled through a sine wave of 0.6 Hz which is very simple to interpolate using cubic splines while the real BLWs are more complex.

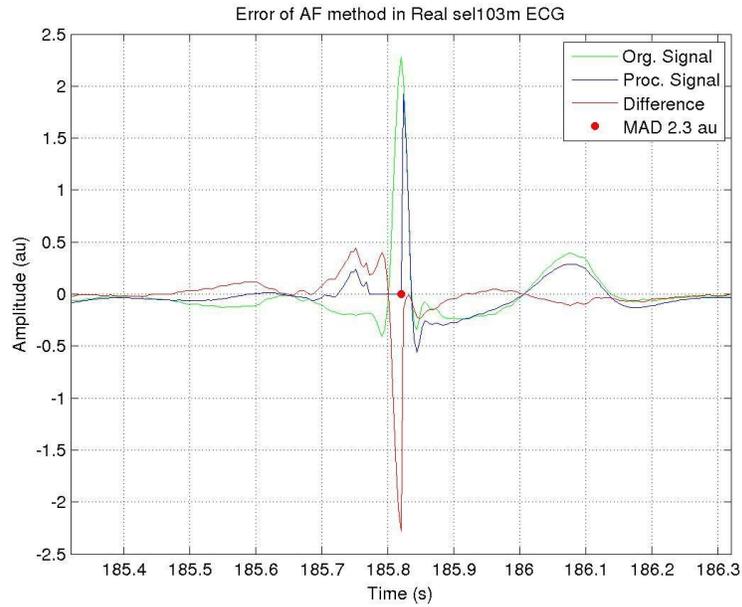

**Figure 13.** Performance of the method based on adaptive filtering on real ECG contaminated with real noise. 1 s segment, centered on the point of greatest distortion, according to the MAD metric. The green trace shows the original signal. The blue trace shows the signal processed by the method. The red trace shows the difference between the 2 analyzed signals.

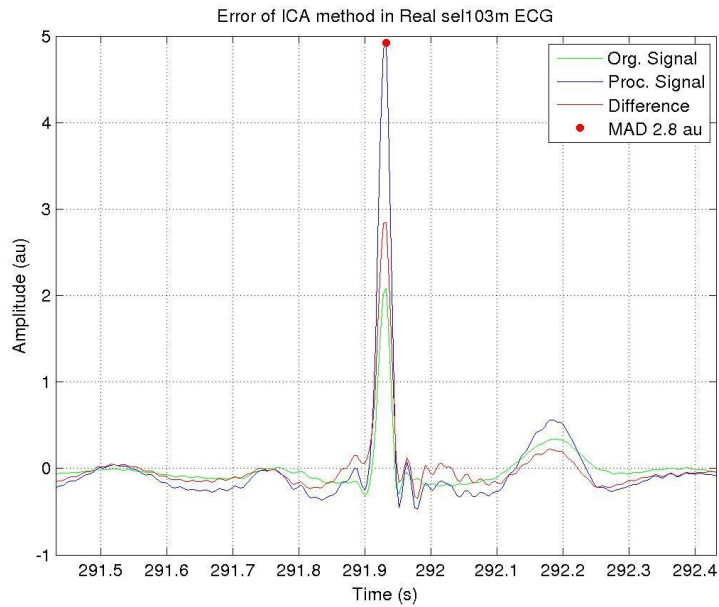

**Figure 14.** Performance of the ICA-based method in real ECG contaminated with real noise. 1 s segment, centered on the point of greatest distortion, according to the MAD metric. The green trace shows the original signal. The blue trace shows the signal processed by the method. The red trace shows the difference between the 2 analyzed signals.

## 5. CONCLUSIONS

In this work was presented a quantitative and qualitative comparison of the performance of the nine most cited methods for the elimination of baseline wanderings, based on the maximum similarity between the input and output signals (i.e. minimum distortion in the ECG signal). The novelty of the study is that, unlike other previous works, the performance of the evaluated methods has been carried out in equal conditions: using identical ECG signals (artificial and real) that are representative of situations of ambulatory monitoring or stress tests, and using three similarity metrics based on distance and not on perform average. The best performance using real signals, according to the three metrics used, was obtained by the method based on digital FIR filter with a cut-off frequency of 0.67 Hz. Considering the low distortion, simple implementation and the low computational cost with respect to other methods, we recommend the use of FIR filter for the elimination of BLW in ECG signals both in embedded devices and in the ECG process using personal computers.